\newif\ifws
\ifws

\documentclass{article}

\usepackage{graphicx}        
\usepackage{xcolor}

\usepackage{hyperref}
\hypersetup{
    colorlinks,
    linkcolor={blue!80!black},
    citecolor={red!75!black},
    urlcolor={blue!80!black}
}

\title{The Emergence of Cognition and Computation: A Physicalistic Perspective}
\author{Karl Svozil \\
        Institute for Theoretical Physics,
Vienna  University of Technology,  \\
Wiedner Hauptstrasse 8-10/136,
1040 Vienna,  Austria
        }

\date{\today}
\begin{document}

\maketitle

\begin{abstract}
A physicalistic argument can support the idea that cognition is an emergent property driven by dissipation. This argument suggests that cognition arises not from any fiat desire to understand the world, but rather because a certain type of cognition promotes dissipation, which is advantageous for agents seeking to increase the dissipation of resources, especially energy, in their favor. In other words, cognitive agents are better equipped to acquire physical resources and means, giving them an advantage in survival and reproduction.
Similarly, the efficient use of computation can also serve as a means of dissipating energy for the computing agent. Efficiency, in this context, is not determined by moral or ethical principles, but rather by the ability to effectively aggregate resources. When used efficiently, computation becomes a powerful tool for dissipating energy and enhancing the survival and reproduction of the agent utilizing it.
\end{abstract}

\else
\documentclass[%
  twocolumn,
 showpacs,
 showkeys,
 preprintnumbers,
 amsmath,amssymb,
 aps,
  pra,
  longbibliography,
 floatfix,
 ]{revtex4-1}

\usepackage{mathptmx}

\usepackage{amssymb,amsthm,amsmath}

\usepackage{tikz}
\usepackage{graphicx}

\usepackage{xcolor}

\usepackage{hyperref}
\hypersetup{
    colorlinks,
    linkcolor={blue},
    citecolor={red!75!black},
    urlcolor={blue}
}

\begin{document}

\title{The Emergence of Cognition and Computation: A Physicalistic Perspective}

\author{Karl Svozil}
\email{svozil@tuwien.ac.at}
\homepage{http://tph.tuwien.ac.at/~svozil}

\affiliation{Institute for Theoretical Physics,
Vienna  University of Technology,
Wiedner Hauptstrasse 8-10/136,
1040 Vienna,  Austria}

\date{\today}

\begin{abstract}
A physicalistic argument can support the idea that cognition is an emergent property driven by dissipation. This argument suggests that cognition arises not from any fiat desire to understand the world, but rather because a certain type of cognition promotes dissipation, which is advantageous for agents seeking to increase the dissipation of resources, especially energy, in their favor. In other words, cognitive agents are better equipped to acquire physical resources and means, giving them an advantage in survival and reproduction.
Similarly, the efficient use of computation can also serve as a means of dissipating energy for the computing agent. Efficiency, in this context, is not determined by moral or ethical principles, but rather by the ability to effectively aggregate resources. When used efficiently, computation becomes a powerful tool for dissipating energy and enhancing the survival and reproduction of the agent utilizing it.
\end{abstract}


\maketitle

\fi


\section{Caveat}

The following ideas are highly speculative. They are intended to stimulate further thought and discussion in the field of cognitive science and artificial intelligence. Readers are advised to approach the following ideas with caution and skepticism, and to consider them as hypotheses to be tested and refined through further research and analysis.

Nevertheless, the ideas presented are grounded in and  based on previous research on cognitive science by scholars such as
Paul Thagard~\cite{thagard_2022,Thagard2019Feb,thagard_2012,Thagard_2010,Thagard2005Feb}, James L. McClelland~\cite{McClelland_2009}, Nikolay Perunov, Robert A. Marsland,
and Jeremy L. England~\cite{England_2013,England_2015,England-PhysRevX.6.021036}, and Daniel C. Dennett~\cite{Dennett1992Oct,Dennett_2005}.
These considerations can be understood as part of a broader context of general questions on the physical aspects of life~\cite{Schwille_2017,te_Brinke_2018,Werlang_2022},
as posed by Erwin Schr\o"dinger~\cite{Schrodinger_1992wil,PHILLIPS2021465}, and self-organization in nonequilibrium systems, as studied by Ilya Prigogine and Gregoire Nicolis~\cite{Nicolis1977May}.
I also draw inspiration from recent advances in Large Language Models (LLMs) of Artificial Intelligence (AI),
such as Generative Pre-trained Transformer (GPT)~\cite{Brown2020May,OpenAI-GTP-4-2023Mar,Anshu2023Mar}.

\section{The Enigma of Existence}

To set the foundation for our discussion on cognition,
it is important to acknowledge that the very existence of the things we seek
to understand---as well as our total lack of comprehension thereof---is the starting point for any meaningful inquiry.
Indeed, the question of existence lies at the heart of many philosophical and scientific inquiries.
It is an enigma that has puzzled scholars and thinkers for centuries. In this section, we explore some of the different perspectives on this issue, ranging from the historical to the subjective.

One of the earliest discussions of existence came from Leibniz~\cite[p.~639]{Leibniz1989}, who famously asked why there is something rather than nothing.
This question remains relevant today, and many philosophers and scientists have weighed in with their own ideas~\cite{sep-nothingness,Rundle-04,gericke2008,gruenbaum-2008,Krauss-2012,Lynds-2012,Bilimoria2012,Goldschmidt2013-GOLTPO-33,carol-witsrtn-2018}.
Wittgenstein argued that
``it is not {\em how}  things are in the world that is the mystical, but {\em that} it exists''~\cite[6.44]{Wittgenstein-TLP}.
In his Freiburg lectures on metaphysics Heidegger similarly posed the {\it ``Angstfrage'':}
why there is something rather than nothing~\cite{Heidegger-1929,Heidegger-1935}.

Despite these questions, some argue that metaphysical questions like the enigma of existence are meaningless.
Wittgenstein famously claimed that the only meaningful statements are those of empirical science~\cite[4.11]{Wittgenstein-TLP},
and that what we cannot speak about we must pass over in silence~\cite[7]{Wittgenstein-TLP}.

However, questions about existence do not disappear by simply ignoring them. Acknowledging our incapacity to fully comprehend ``existence'' enables us to explore the limits of our understanding and challenge our assumptions about the world. By meditating on the fact of our own existence, we can recognize its incomprehensibility and gain a better understanding of our limitations. This approach may also lead to a more humble and open-minded interpretation of religious experiences.

\section{Distinctions, and interfaces: understanding the partitioning of existence}

Once we acknowledge the existence of the universe, we can further assume that this universe is partitioned into distinct parts~\cite{Spencer-Brown2008Sep}, each defined by its own borders.
These borders act as interfaces between the parts and allow for mutual interconnection.

An interface can be defined as the boundary or surface that separates two distinct regions or phases of a system~\cite{Diebner2000Jul}.
By recognizing the partitioning of the universe and the interconnectedness of its parts through interfaces, we can begin to explore the dynamics of the system as a whole.

\section{Maximizing diffusion for the acquisition of resources}

Diffusion refers to the overall movement of a resource, substance, entity or category, be it atoms, molecules, or energy, from a region of high concentration to a region of lower concentration. This movement is driven by a gradient in Gibbs free energy or chemical potential, which propels the substance down its concentration gradient.

Diffusion is a ubiquitous phenomenon that occurs on different scales and in various contexts.
It is observed in heat conduction in fluids, nuclear reactor operation, perfume spreading in a room, ions crossing a membrane, and energy flow in organisms, tools, and societies. Understanding the mechanisms and processes that drive diffusion is essential in fields such as chemistry, physics, and biology, as it plays a fundamental role in the behavior of materials and systems at different scales.

Diffusion is closely related to the Second Law of Thermodynamics, which states that the total entropy (or disorder) of a closed system can only increase or remain constant over time. Diffusion occurs spontaneously and leads to an increase in entropy, as the movement of particles from a region of high concentration to a region of low concentration leads to a more uniform distribution of particles.
This increase in entropy is a result of the random motion of particles, which eventually leads to their dispersion throughout the available space.
Therefore, diffusion is an example of a natural process that is consistent with the Second Law of Thermodynamics~\cite{Myrvold2011237}, and it can be used to illustrate the concept
of entropy and its relation to the spontaneous movement of matter and energy in physical systems~\cite{Fick_1855,Fick_1855e,Philibert2005}.

Therefore, once different parts of the universe have varying concentrations of entities such as energy, the second law implies that diffusion occurs through interfaces connecting these parts.

The rate of diffusion, which refers to the transfer of resources, substances, entities or categories (e.g. atoms, ions, molecules, energy) from one part of the universe to another, is influenced not only by the concentration difference, but also by the interface's ability to transport and transfer these goodies~\cite{Fick_1855,Fick_1855e,Philibert2005}.

In situations where there is a highly concentrated reservoir, and two or more less concentrated reservoirs, we can conceive of a ``competition'' between the parts or reservoirs with lower concentration to draw resources from the highly concentrated reservoir. Once diffusion has taken place, the parts or reservoirs that were previously less concentrated will end up with amounts of resources proportional to the capacity of their interfaces.

\section{Cognition: a key component in the evolutionary toolbox}


\subsection{Main thesis}

Our main hypothesis is that cognition, and its derivative computation, has evolved as a means to increase the capacity of the interface
between an organism and its environment.
Through cognition, organisms (or agents and organizations) can physically alter the interface to their advantage, allowing them to extract and utilize resources more efficiently.
As a result, the capacity to process information and make decisions becomes a critical factor in determining an organism's survival and reproductive success.
This also encompasses interface extensions and expansions that could arise from accessing ``distant reservoirs'' like oil or uranium deposits.

Pointedly stated, a species with higher cognitive and technological capabilities can extract more resources from its surroundings.
This presents a competitive advantage in the struggle for resources, allowing individuals, tribes, groups, societies, and species to flourish.
Thus, we propose that cognition and computation have developed not through an arbitrary ``ad hoc'' or fiat process,   but rather as a tool
to ``conquer abundance''~\cite{Feyerabend2001May} and outcompete and outsmart other organisms for resources.

Overall, our hypothesis suggests that cognition has played a crucial role in the evolutionary success of organisms,
by enabling them to adapt to changing environments and extract resources more efficiently.
By understanding the mechanisms behind this evolutionary advantage, we can gain valuable insights into the nature of cognition and its role in shaping the biology of living organisms.

\subsection{Formalization}

As already discussed earlier, suppose that there are two distinct regions in space with varying temperatures or energy densities. The reason for this difference may be attributed to fluctuations or initial values, but we will not delve into this further.
Now, imagine that there is a medium that connects these two regions, which could be empty space, a material structure, or an agent that allows for physical dissipative flows to occur between them.
In accordance with the second law of thermodynamics, energy will naturally flow from the hotter region to the colder region through this interface.
This is a fundamental physical process that occurs regardless of external influence~\cite{Fick_1855,Fick_1855e,Philibert2005}.

In its simplest form---two  ``infinite'' reservoirs, one with a ``high''  concentration  of some resource,
and another one with a ``lower'' concentration of that resource,
and a boundary or interface between them---the flow of that resource  $\varphi $---that is, the change of the resource $\Delta \varphi $ per time interval  $\Delta t$
as a function of time $t$---can
be modelled (that is, assumed) to be constant in time:
\begin{equation}
\lim_{\Delta t \rightarrow 0}\frac{\Delta \varphi }{\Delta t} =  \frac{d}{dt} \varphi = a.
\end{equation}
Direct integration
\begin{equation}
\int_{\varphi(0)}^{\varphi(T)} d \varphi = \varphi(T) - \varphi(0) = \int_0^T a dt = a T
\end{equation}
yields the sum total
\begin{equation}
\varphi(T)= \varphi(0)+ aT
\end{equation}
of
the resource $\varphi$ available in the lower concentration reservoire at time $T$.
The amount of resource $\varphi(T) - \varphi(0) = a T$ ``drawn'' from the higher concentration reservoire is linear in time, and proportional to $a$.
Clearly, a  higher throughput rate with higher coefficient $a$ indicates linear higher transfers of resources.

Let us turn our attention to the interface between the two regions.
Specifically, we will consider a variety of interfaces and assess their relative efficiency or ``fitness.''
This concept takes us into the realm of evolutionary biology.
Assuming all other factors are equal, the interface with the highest energy throughput rate (in the earlier example, $a$) will dominate the dissipation process,
effectively grabbing the largest share of available energy.

To find the most optimal interfaces, we can facilitate the process through random mutation and spontaneous exchanges of the genotype,
allowing exploration of the abstract space of possible interface states and configurations. A useful algorithmic
expression for this process is genetic algorithms~\cite{Srinivas_1994,Whitley_2012,Katoch_2020}.
This technique involves mimicking the principles of biological evolution by using natural selection,
mutation, and recombination of genetic information to optimize the interface's fitness.
By iterating through generations of interfaces, genetic algorithms can quickly and effectively identify the most efficient and robust solutions.

The situation becomes even more dynamic when the relative magnitude of the different processes changes over time.
In particular, if a highly efficient process can self-replicate or perform recursively~\cite{Smullyan1993-SMURTF,book:486992}, for instance,
utilize feedback-loops.
One example is a regime dominated by the ``Matthew effect''~\cite{merton-68} of compound resources.
This means that the population of the most potent interfaces will increase at a rate of compound interest, which is essentially exponential.
Growth rates may appear linear initially (and therefore sustainable), but they will accelerate until all available energy is distributed,
or other limiting factors come into play.

In the model discussed earlier the Matthew effect of compound resources can be formalized by assuming that (parts of) the resources drawn from the high concentration reservoire
are ``re-invested'' into extension of the interface, so that, say, the dependency (aka increase for positive flow) of the diffusion becomes linear with the resources drawn,
that is,
\begin{equation}
\frac{d}{dt} \varphi(t) = b \varphi(t)\text{, or } \varphi'(t)-b  \varphi(t)=0.
\end{equation}
This is a ``first order'' Fuchsian equation~\cite{larson-edwards-calculus}  which, for instance, can be solved with the
{\em Frobenius method}~\cite{arfken05}.
The solution is
\begin{equation}
\varphi(t) = \varphi(0) \exp(b  t),
\end{equation}
which indicates an exponention growth in dissipation and amount of resource as long as there are no further constraints.

If we identify certain interfaces with biological entities, we arrive at a type of biological evolution driven by physical processes, specifically energy dissipation.
This idea has been explored in recent research by Jeremy L. England, Nikolay Perunov, and Robert A. Marsland~\cite{England_2013,England_2015,England-PhysRevX.6.021036}.

The emergence of computation is relevant to this picture, and its connection is relatively straightforward if we continue to explore this speculative path.
Systems capable of computation can serve as, or even construct and produce, interfaces
that are better suited for energy dissipation than those without algorithmic abilities.
Through the process of mutation and trial-and-error driven by random walks through roaming configurations and state space, the universe,
self-reproducing agents, and units have learned to compute. This process, in essence, is a scenario for the emergence of mathematics and universal computation.
Once universal computation is achieved ``the sky'' or rather recursion theory is the limit~\cite{Smullyan1993-SMURTF,book:486992}.

By developing computational abilities, systems can optimize their interfaces for energy dissipation,
leading to more efficient and effective energy transfer and dissipation.
Therefore, computation is an essential factor in the evolution of complex systems, providing them with a powerful tool for adapting to their environment and improving their fitness over time.

Taking this speculation further, one could suggest that self-awareness and consciousness emerged in a similar way, driven by the imperative to dissipate
or ``use and realize and burn'' to one's own advantage as much energy or as many resources as possible.
In other words, the ability to perceive and represent the self may have evolved as a means of optimizing energy transfer and dissipation.

Consciousness, then, could be seen as a product of the same evolutionary forces that led to the emergence of more efficient interfaces for energy dissipation.
As systems became more complex and better able to manipulate their environment,
 the ability to perceive and understand their own actions and their consequences became increasingly important for optimizing energy use and dissipation.
This process could have culminated in the emergence of computation aka recursion, self-referentiality, self-awareness and consciousness,
which are now seen as not only defining features but also
limitations of human cognition and experience~\cite{chaitin3,calude:02,Yanofsky2016,Yanofsky-BSL:9051621,Yanofsky-object}.

As humorously illustrated in Fig.~\ref{2017-ptuc-f}, computation, mathematics, and the human mind may have emerged as mere tools for facilitating optimal heat exchange.
They may have evolved as a means of optimizing energy transfer and dissipation, and continue to be driven by this imperative.
In other words, the universe seeks to better understand and perceive itself in order to achieve optimized ``self-digestion,'' or efficient dissipation of energy.
Therefore, the emergence and evolution of complex systems, including those capable of computation and consciousness,
can be seen as a natural consequence of the universe's drive to optimize energy use and dissipation.
\ifws
\begin{figure}
\begin{center}
\includegraphics[width=5cm,angle=0]{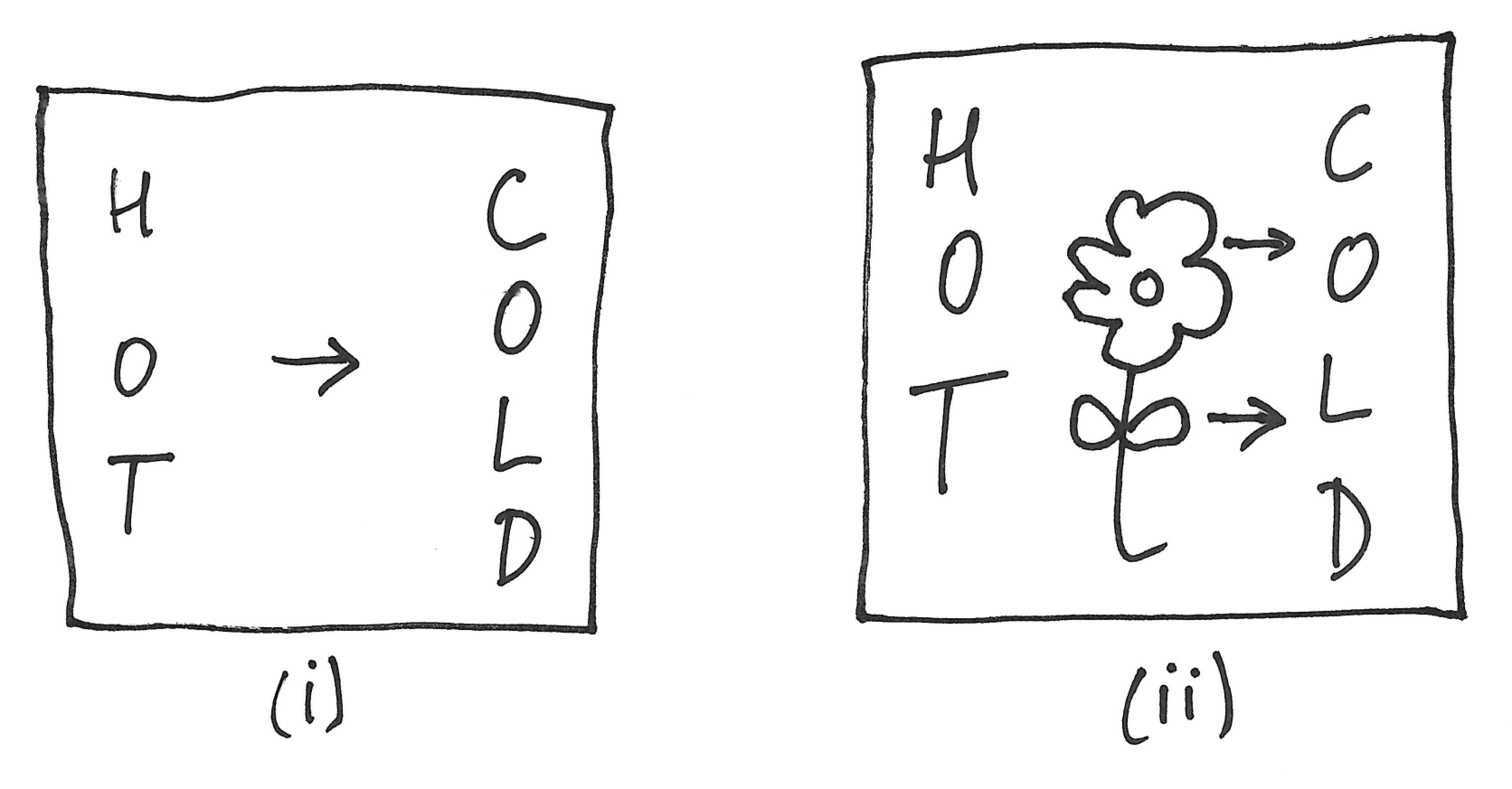}
\hspace{3mm}
\includegraphics[width=5cm,angle=0]{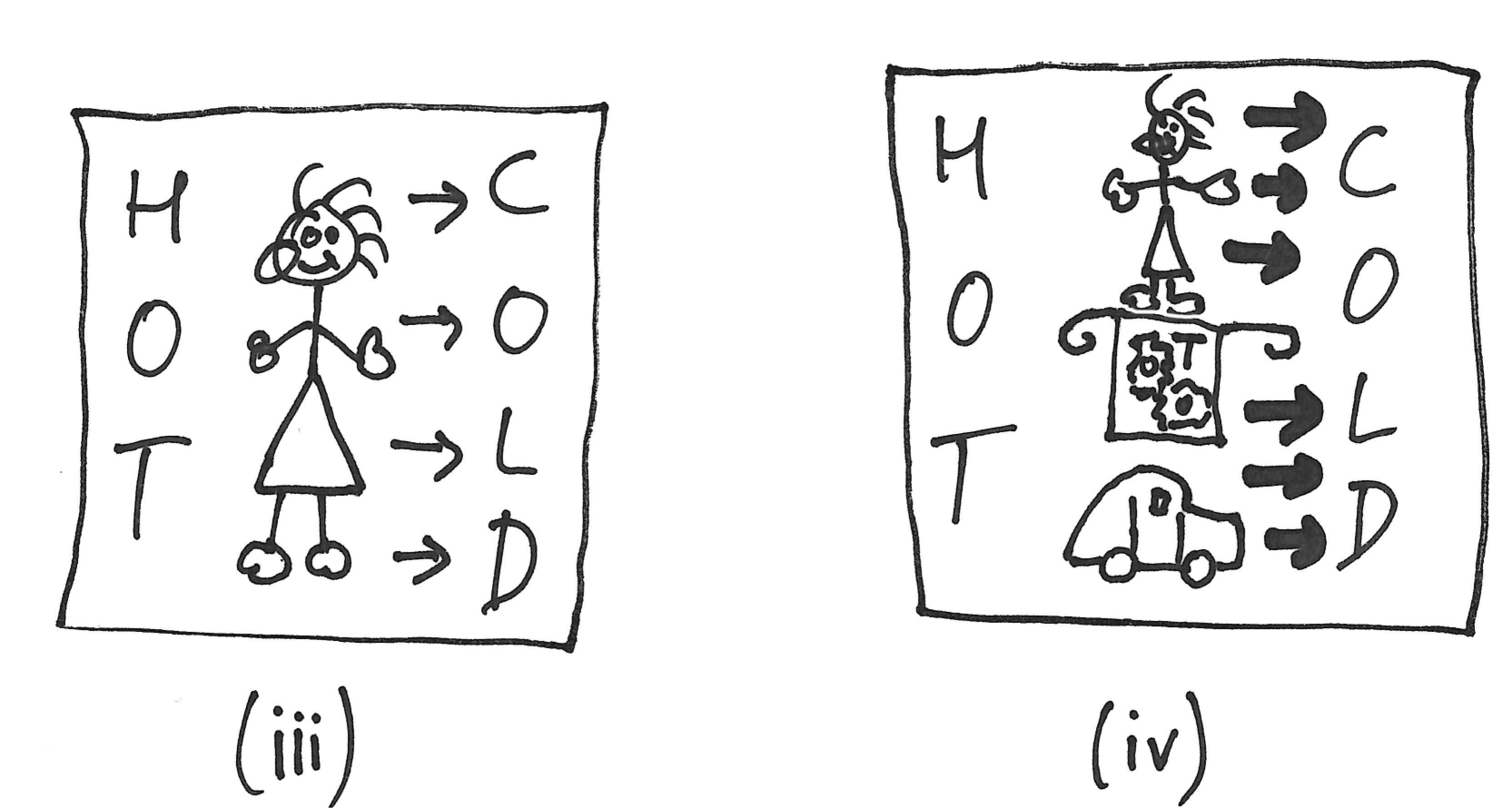}
\end{center}
\caption{
Evolution of species and computation, driven by the second law of thermodynamics (as inspired by England's {\it et al} approach~\cite{England_2015,England-PhysRevX.6.021036}):
(i) interface between hot and cold regions is empty space;
(ii) plant interface capable of more dissipation than empty space;
(iii) animals and, in particular, humans (drawn political correctly) present interfaces with improved (over plants and emptyness) energy dissipation;
(iv) humans equipped with engines and universal computation capacities (indicated by ``T'' for ``universal Turing machine'') can consume even more energy than standalone.
\label{2017-ptuc-f}
}
\end{figure}
\else
\begin{figure*}
\begin{center}
\includegraphics[width=7cm,angle=0]{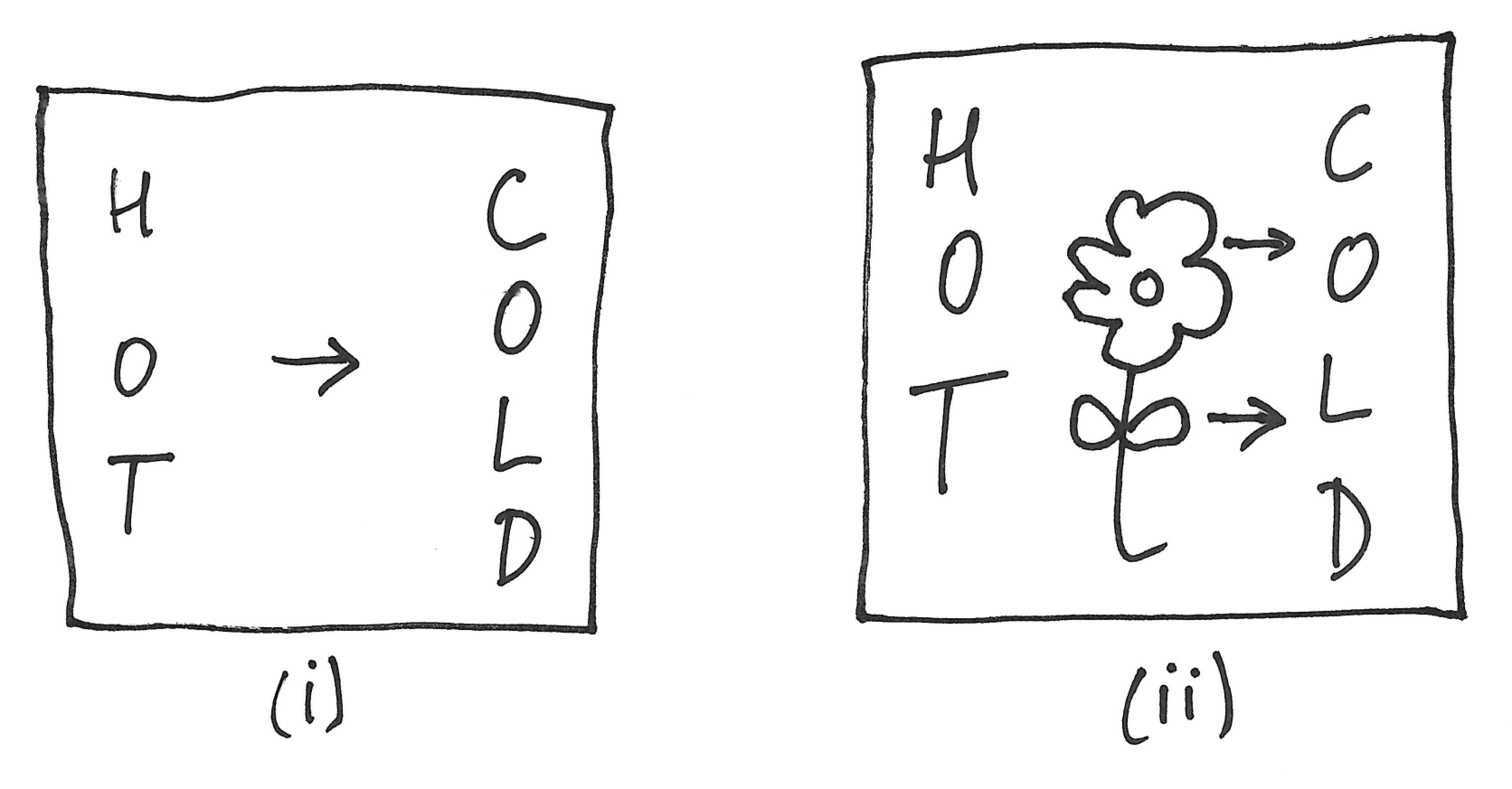}
\hspace{3mm}
\includegraphics[width=7cm,angle=0]{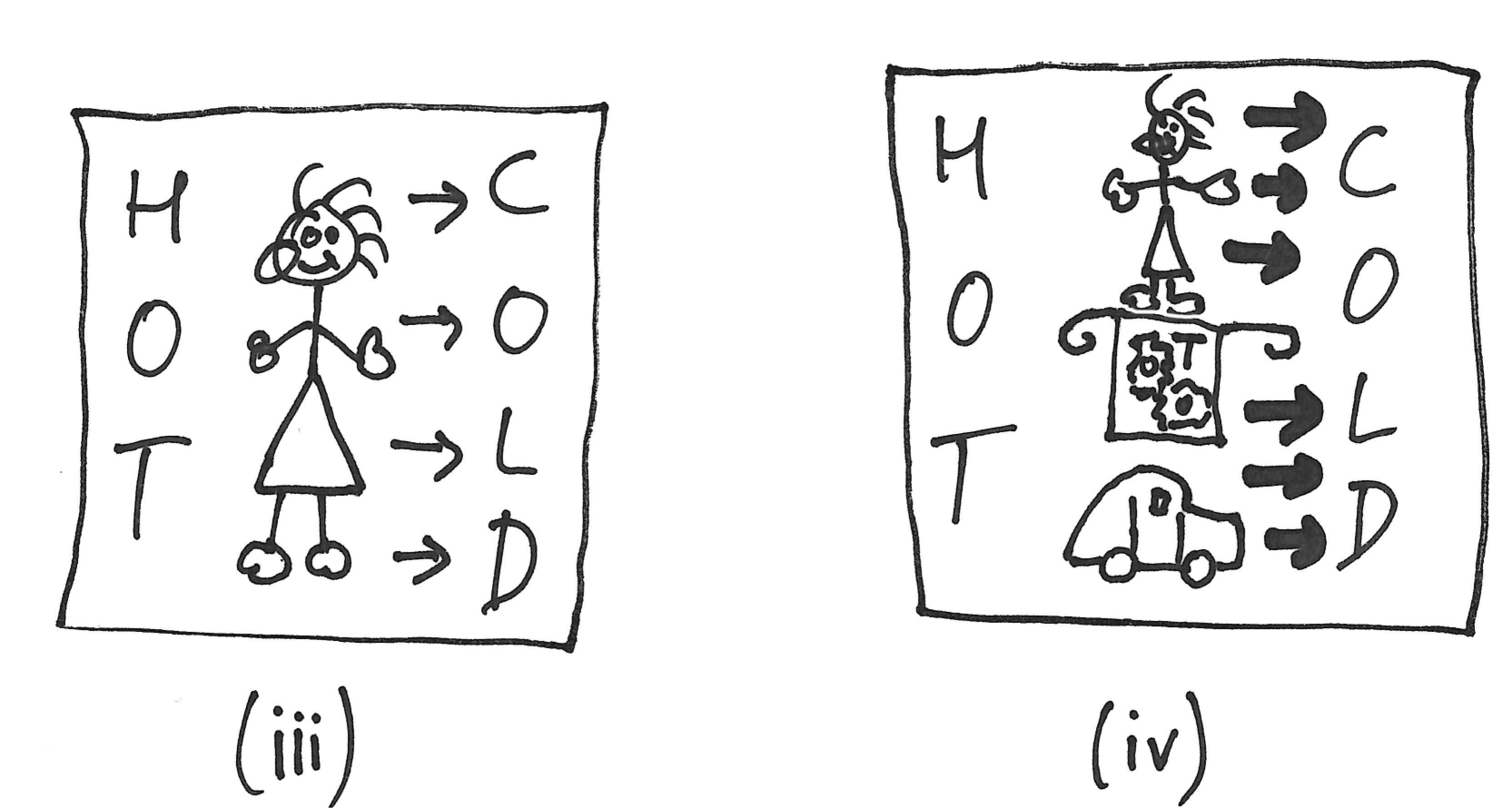}
\end{center}
\caption{
Evolution of species and computation, driven by the second law of thermodynamics (as inspired by England's {\it et al} approach~\cite{England_2015,England-PhysRevX.6.021036}):
(i) interface between hot and cold regions is empty space;
(ii) plant interface capable of more dissipation than empty space;
(iii) animals and, in particular, humans (drawn political correctly) present interfaces with improved (over plants and emptyness) energy dissipation;
(iv) humans with engines and universal computation capacities (indicated by ``T'' for ``universal Turing machine'') can consume even more energy than standalone.
\label{2017-ptuc-f}
}
\end{figure*}
\fi

\subsection{Discussion}

This perspective may seem rather bleak or even dystopian, as it suggests that the evolution of species, consciousness, mathematics,
and computation all arise as means to consumption and access more and more resources such as food or energy than would occur without these emergent systems.
Ethical and even theological questions naturally arise from this view.

However, we must recognize that even in a universe of primordial chaos, where the laws of nature may be fundamentally probabilistic and subject to deviations from their expected form on small scales, the emergence of order and principles of social conditioning may still occur.
As Egon von Schweidler suggested in 1905 with regards to radioactive decay~\cite{schweidler-1905},
all of our natural laws be subject to probabilistic fluctuations.
This idea, originally proposed by Exner~\cite{Exner-1908,Hanley-1979} and reviewed by  Schr\"odinger~\cite{schrodinger-1929,book:16081},
suggests that the very foundations of our universe may be subject to deviations from expected behavior.

In this context, it is possible that the immanent emergence of gods, law, and order occurred as a means of discourse as well as of social conditioning,
particularly as societies formed and secular and religious powers underwent a symbiotic relationship.
Ultimately, questions of ethics and divinity remain pertinent---recall the earlier contemplation on ``existence''---but
it is important to recognize that these may be shaped by, and subject to,
the very same forces that drive the evolution of complex systems in the universe.


\section{Can language models shed light on the nature of self-awareness?}

In this section, I will present an argument in favor of using LLMs as a suitable model for understanding the world.
To begin, let us revisit Heinrich Hertz's views on physical model building,
as stated in the introduction of his work on classical mechanics~\cite{hertz-94,hertz-94e} (my emphasis):
\begin{quote}
``The most direct, and in a sense the most important, problem
which our conscious knowledge of nature should enable us to
solve is the anticipation of future events, so that we may
arrange our present affairs in accordance with such anticipation.
As a basis for the solution of this problem we always
make use of our knowledge of events which have already
occurred, obtained by chance observation or by prearranged
experiment. In endeavouring thus to draw inferences as to
the future from the past, we always adopt the following process.
We form for ourselves images or symbols of external objects;
and the form which we give them is such that {\em the necessary
consequents of the images in thought are always the images of
the necessary consequents in nature of the things pictured.} In
order that this requirement may be satisfied, there must be a
certain conformity between nature and our thought. Experience
teaches us that the requirement can be satisfied, and hence that
such a conformity does in fact exist.
$\ldots$
For our purpose it is not
necessary that they should be in conformity with the things in
any other respect whatever. As a matter of fact, we do not
know, nor have we any means of knowing, whether our conceptions
of things are in conformity with them in any other
than this one fundamental respect.

The images which we may form of things are not determined
without ambiguity by the requirement that {\em the consequents
of the images must be the images of the consequents.}''
\end{quote}

Hertz's principle that ``the consequents of the images must be the images of the consequents'' is highly applicable to the LLM context, as it directly corresponds to the way LLMs operate. Specifically, pre-training, which involves training an LLM on a large corpus of unlabeled text data to acquire comprehensive language knowledge and representations, can be seen as a manifestation of this principle.
Two common methods of pre-training~\cite{Brown2020May,OpenAI-GTP-4-2023Mar,Anshu2023Mar} are Masked Language Modeling and Autoregressive Language Modeling.

Masked language modeling (MLM)~\cite{Mialon2023Feb} is a powerful technique in which certain tokens within the input text are randomly masked, and the LLM is then trained to predict these masked tokens by analyzing the surrounding context. A token is a sequence of characters that represents a single unit of meaning in a text sequence, such as a word, which is obtained through the process of tokenization, the breakdown of text into individual tokens. By using MLM, LLMs can learn bidirectional context and capture long-range dependencies between words, resulting in more accurate predictions and a better understanding of the text.

Autoregressive language modeling (ALM)~\cite{Mialon2023Feb,Liu2022Oct} is another powerful technique that trains an LLM to predict the next token in a text sequence given the preceding tokens. This method is extensively employed in GPT and its variants, such as GPT-2, GPT-3, and others. ALM allows LLMs to learn causal relationships between words and generate fluent text, resulting in more natural-sounding and coherent language generation. By sequentially predicting the next token, ALM enables LLMs to generate lengthy text passages that are grammatically and semantically correct,
making it an invaluable tool for a range of natural language processing applications.

In a very crude way, tokens of LLM's can be compared to Hertz's images, and the way to compound scientific knwowledge is by adapting LLM's to token predictions.

\section{Exploring Almost Quantum-Like Representations for LLMs}

Vector representations in LLMs like refer to the process of converting tokens such as (parts of) words or sequences of words into
high-dimensional numerical vectors---from 768 up to several thousand dimensions---that can be processed by the model's neural network.
These vectors are typically dense, meaning they contain a large number of non-zero values and
are designed to capture the semantic and syntactic relationships between words and phrases.
Depending on the specific configuration of the model, the components or coordinates of this vectors are floating-point numbers of either 16-bit or 32-bit precision.

LLMs  are based on vector representations.
A vector representation is a way of encoding a word or a symbol as a numerical vector,
usually with a fixed number of dimensions.
Vector representations allow language models to capture semantic and syntactic similarities between words or symbols,
and to perform mathematical operations on them.

The ``proximity'' of vectors can be formalized by metric such as the standard Euclidean metric measuring the angle between vectors.
The processing of vectors by a model's neural network can often be represented mathematically as a series of matrix multiplications and nonlinear transformations.

In a typical neural network architecture, the vector representations of input data (such as text) are fed into the network as input to the first layer of the network. Each layer of the network then applies a series of matrix multiplications and nonlinear transformations to the input vector, transforming it into a new vector that captures more complex features of the data.

This process of matrix multiplication and nonlinear transformation is often referred to as a forward pass through the network, and it can be represented mathematically as a series of matrix-vector multiplications, followed by the application of a nonlinear activation function.

The parameters of the network, including the weights and biases of each layer, are typically stored as matrices and vectors, and are updated through a process of backpropagation and gradient descent during training.

Overall, the processing of vectors by a neural network can be represented mathematically as a series of matrix operations, making it possible to analyze and optimize the network using techniques from linear algebra and calculus.

This ``almost'' (due to the presence of nonlinearity) Hilbert space-like formalization of knowledge processing and prediction in LLMs exhibits striking similarities to the quantum evolution of a quantum state. By modeling the vector representations of words and phrases as quantum states in an ``almost'' Hilbert space, LLMs can leverage the mathematical framework of quantum mechanics to perform computations and predictions on natural language data.

The use of an ``almost'' Hilbert space to model LLMs is inspired by the concept of quantum-like behavior observed in cognitive systems, in which the dynamics of information processing exhibit patterns similar to those of quantum mechanics. This approach enables LLMs to capture the complex and subtle relationships between words and phrases in a way that is mathematically rigorous and computationally efficient.

Furthermore, the use of an ``almost'' Hilbert space allows LLMs to model the inherent uncertainty and ambiguity of natural language, much like quantum mechanics can model the inherent uncertainty of physical systems. This enables LLMs to generate more nuanced and contextually appropriate responses to natural language input.

Overall, the ``almost'' Hilbert space-like formalization of knowledge processing and prediction in LLMs represents an innovative and promising approach to natural language processing that draws on insights from quantum mechanics and cognitive science.

The versatility of (almost) Hilbert space representations in LLMs, which are used to encode tokens and processes, reveals potential connections to the usefulness, utility, and even necessity of the quantum formalism. The Hilbert space representations are a mathematical structure that describes a quantum system's state, including its observable properties and possible measurements.
This could provide connections between quantum mechanics and natural language processing, which could lead to significant advancements in comprehending
and ultimately extending both fields.

\section{Some afterthoughts}

The concept of a physical foundation for consciousness, as previously discussed, follows
in the spirit of Landauer's assertion that ``information is physical''~\index{landauer}, and extends it to ``consciousness is physical.''
This perspective offers an explanation without invoking a divine entity or resorting to a ``god of the gaps''~\cite[Chapters~III.12-15]{frank,franke},
although it does not account for the initial boot-up of the universe.

I am aware that readers might object my pretense to reduce or relate their behavior and therefore their cognitive capacities to LLMs.
However, there may be some empirical evidence that at least part of the human cognition is steered by LLMs, although we seem to have the capacity to inhibit and decide agains this
continuous flow of motions.
In the Libet experiment~\cite{libet}, participants were asked to perform a simple task, such as pressing a button, while their brain activity was being monitored. The experiment found that there was a detectable buildup of electrical activity in the brain's motor cortex before the participants reported the conscious decision to move,
suggesting that the decision to move may have already been made at an unconscious level.

Another difference of human cognition with respect to LLMs might be strong mechanisms of censorship, as well as for rewards.
The neural mechanisms that underlie reward processing involve the release of certain neurotransmitters,
such as dopamine, in specific brain regions, such as the mesolimbic pathway.
In the 1950s, James Olds and Peter Milner conducted an experiment where rats were given direct brain stimulation through implanted electrodes.
The rats would press a lever to activate the pleasure center in their brain and would do so repeatedly, even to the point of ignoring food, water, and even their own offspring. Some rats would self-stimulate up to 2000 times per hour for 24 hours, to the exclusion of all other activities,
and had to be disconnected from the apparatus to prevent death by self-starvation~\cite{Olds_1954,Milner_1989,Moan_1972,Portenoy_1986,Linden2012Apr,Linden2011Sep,Lieberman2018Aug}.

The Libet experiment provides evidence that a significant~\cite{huxley} portion of cognitive processes is subconscious,
 meaning they occur below the threshold of conscious awareness but still impact our actions and feelings.
This suggests that our conscious experience is only the tip of the iceberg when it comes to the inner workings of our minds.

Moreover, discussing consciousness, feelings, and awareness, the ``sentient I,''
can be particularly challenging, as these concepts are difficult to define and operationalize.
Testing for them is also notoriously elusive, and it can be difficult to determine whether an LLM or a human individual is truly conscious.
As Descartes famously noted in his Meditations~\cite{descartes-meditation}, the only thing one can be certain of is one's own existence (``Cogito, ergo sum'').

One of the main objectives of future research will be the development of a rigorous theoretical framework for the various notions and hypotheses that we have discussed in this paper.
Specifically, we would like to give a precise definition of what we mean by ``emergence''~\cite{anderson:73,Wei2022Jun} in the context of dissipative systems,
and how this concept is related to other important notions such as complexity, information, and functionality.
Furthermore, we would need to test the validity and applicability of our main hypothesis,
which proposes that ``emergence by optimizing the dissipation of energy'' is a general principle
that can account for the origin and evolution of lifelike behaviors and other functionalities,
including cognition and universal computation, in nonequilibrium systems.
We also plan to explore the limitations and challenges of this hypothesis,
such as the effects of (thermal) noise and finite (energy) resources, as well as the interactions with other systems or observers,
on the emergent properties and processes.

\bibliography{svozil}
\ifws

\bibliographystyle{spmpsci}

\else
 \bibliographystyle{apsrev}

\fi

\end{document}